%% file: main.tex
\documentclass[10pt,twocolumn,letterpaper]{article}

\usepackage{cvpr}              

\input{preamble}

\definecolor{cvprblue}{rgb}{0.21,0.49,0.74}
\usepackage[pagebackref,breaklinks,colorlinks,citecolor=cvprblue]{hyperref}
\usepackage{multirow}

\def\paperID{10} 
\def\confName{CVPR}
\def\confYear{2024}

\title{MIPI 2024 Challenge on Demosaic for Hybridevs Camera: Methods and Results}

\author{
\textbf{Challenge and Workshop Organizers} \\
Yaqi Wu \quad Zhihao Fan \quad Xiaofeng Chu \quad Jimmy S. Ren \quad 
Xiaoming Li \quad Zongsheng Yue \\ Chongyi Li \quad Shangcheng Zhou \quad Ruicheng Feng \quad Yuekun Dai \quad Peiqing Yang \quad Chen Change Loy \\ 
\\
\textbf{Challenge Participants} \\
Senyan Xu \quad Zhijing Sun \quad Jiaying Zhu \quad Yurui Zhu \quad Xueyang Fu \quad Zheng-Jun Zha \\ 
Jun Cao \quad Cheng Li \quad Shu Chen \quad Liang Ma \quad 
Shiyang Zhou \quad Haijin Zeng \quad Kai Feng \\ Yongyong Chen \quad Jingyong Su \quad 
Xianyu Guan \quad Hongyuan Yu \quad Cheng Wan \quad Jiamin Lin \\ Binnan Han \quad Yajun Zou \quad Zhuoyuan Wu \quad Yuan Huang \quad Yongsheng Yu \quad Daoan Zhang \\ Jizhe Li \quad Xuanwu Yin \quad Kunlong Zuo \quad 
Yunfan Lu \quad Yijie Xu \quad Wenzong Ma \\ Weiyu Guo \quad Hui Xiong \quad
Wei Yu \quad Bingchun Luo \quad 
Sabari Nathan \quad Priya Kansal
}

\begin{document}
\maketitle
\input{sec/0_abstract}    
\input{sec/1_intro}
\input{sec/2_track}
\input{sec/3_result}
\input{sec/4_methods}

\input{sec/5_conclusions}
{
    \small
    \bibliographystyle{ieeenat_fullname}
    \bibliography{main}
}

\appendix

\input{sec/6_teams}


\end{document}

%% file: preamble.tex
\usepackage[dvipsnames]{xcolor}


%% file: sec/0_abstract.tex
\begin{abstract}

The rising demand for computational photography on mobile devices drives the development of advanced image sensors and algorithms for camera systems. However, the lack of opportunities for in-depth exchange between industry and academia is constraining the development of Mobile Intelligent Photography and Imaging (MIPI). Building on the successes of the prior MIPI Workshops at ECCV 2022 and CVPR 2023, we are pleased to introduce our third MIPI challenge in conjunction with CVPR 2024, which includes three tracks focusing on novel image sensors and imaging algorithms. In this paper, we summarize and review the Demosaic for the HybridEVS Camera track on MIPI 2024. A total of 110 participants from both industrial and academic backgrounds contributed many valuable solutions to address the difficulty of the restoration of HybridEVS's raw data, thus raising the reconstructed performance to a new height. This paper gives a comprehensive description and analysis of all solutions developed during this challenge. More detailed information about this challenge is available at \href{https://mipi-challenge.org/MIPI2024/}{https://mipi-challenge.org/MIPI2024}.

\end{abstract}

%% file: sec/1_intro.tex
\section{Introduction}
\label{sec:intro}

\begin{figure}[!ht]
\centering
\vspace{-3pt}
\includegraphics[width=0.48\textwidth]{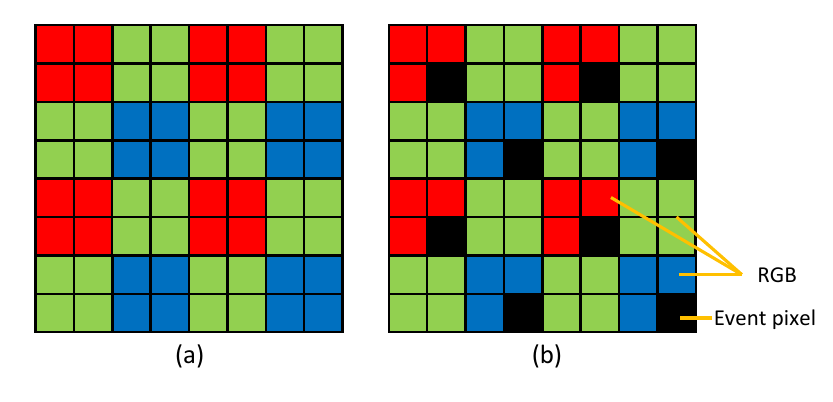}
\caption{(a) Quad Bayer pattern, (b) HybridEVS pattern.}
\label{fig:pattern}
\setlength{\belowcaptionskip}{0pt plus 3pt minus 2pt}
\vspace{-10pt}
\end{figure}

\begin{figure*}[!ht]
\centering
\includegraphics[width=0.9\textwidth]{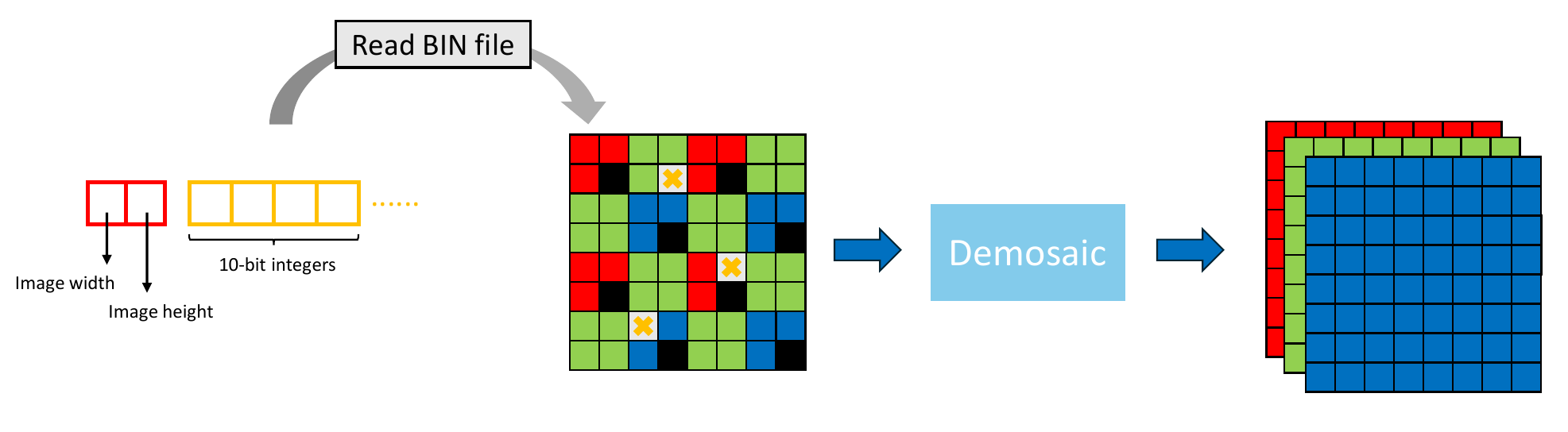}
\caption{The Demosaic for the HybridEVS Camera aims to reconstruct HybridEVS data into a high-quality RGB result with the same resolution. This process involves passing the data through a demosaic module, which corrects defects and event pixels and reconstructs a three-channel RGB image of matching resolution.}
\label{fig:simple_demosaic}
\setlength{\belowcaptionskip}{0pt plus 3pt minus 2pt}
\end{figure*}
%

%
Quad Bayer Color Filter Array (CFA) is a popular CFA pattern widely used in smartphone cameras such as the Galaxy S20 FE and Redmi Note8 Pro (Figure~\ref{fig:pattern}~(a)). Quad Bayer CFA differs from the traditional Bayer CFA by using 2$\times$2 cells of identical color filters. By utilizing demosaic technics, it can acquire high-resolution images with good image quality. Moreover, this design ensures exceptional low-light performance through a 2$\times$2 binning operation. But in existing ISP-related research, exploration of Quad Bayer CFA is very limited, with most pipelines concentrating on Bayer
CFA~\cite{kim2019high, hseih2015new, yonemoto2003principles}.
Event Vision Sensors (EVS) determine, at the pixel level, whether a temporal contrast change beyond a predefined threshold is detected~\cite{gallego2020event,son2017640}. Compared to CMOS image sensors (CIS), this new modality inherently provides data-compression functionality and hence, enables high-speed, low-latency data capture while operating at low power. EVS has tremendous application potential in object tracking, 3D detection, and slow-motion.

Hybrid Event-based Vision Sensor (HybridEVS)~\cite{kodama20231} is a novel hybrid sensor formed by combining Quad Bayer CFA with Event-based Vision techniques. As shown in Figure~\ref{fig:pattern}~(b), within the 4x4 block, two event pixels are used to capture event signals, while the remaining pixels are utilized to obtain color information. Owing to the inability of event pixels to capture color and texture information, demosaicing tasks become more challenging for HybridEVS. 
%
Moreover, due to pixel flaws caused by the sensor's manufacturing process,
defect pixels may occasionally arise, characterized by significantly divergent pixel values from those of unaffected pixels.
%

Given the presence of event pixels and defect pixels, the Demosaic for HybridEVS Camera has become increasingly challenging, with very limited related academic research available. Therefore, we are organizing this competition with the overarching aim of cultivating innovative solutions to elevate the related research level of this task to a new height.
%
We hold this challenge in conjunction with the third MIPI Challenge which will be held on CVPR@2024. Similar to the previous MIPI challenge~\cite{sun2023mipi,sun2023mipi2,dai2023mipi,zhu2023mipi,yang2022mipi}, we are seeking an efficient and high-performance image restoration algorithm to handle the HybridEVS camera demosaic task. MIPI 2024 mainly consists of three competition tracks focusing on the following tasks:

\begin{itemize}
    \item \textbf{Few-shot RAW Image Denoising} is geared towards training neural networks for raw image denoising in the scenarios where paired data is limited.
    \item \textbf{Demosaic for HybridEVS Camera} is to reconstruct HybridEVS's raw data that contains event pixels and defect pixels into RGB images.
    \item \textbf{Nighttime Flare Removal} is to improve nighttime image
    quality by removing lens flare effects.
\end{itemize}

%% file: sec/2_track.tex
\section{MIPI 2024 Demosaic for Hybridevs Camera}
\label{sec:track}

To facilitate the development of efficient and high-performance demosaic solutions, we provide a high-quality dataset to be used for training and testing and a set of evaluation metrics that can measure the performance of developed solutions.
This challenge aims to advance research on demosaic for HybridEVS camera.

\subsection{Problem Definition}
 As illustrated in Figure~\ref{fig:simple_demosaic}, the Demosaic for HybridEVS Camera is dedicated to reconstructing the HybridEVS input data to a promising RGB result. 
Due to manufacturing defects, HybridEVS data may contain defect pixels whose actual values deviate significantly from the ideal values. 
Additionally, the presence of event pixels, essential for capturing motion information, poses challenges to the reconstruction task.
Given a HybridEVS input \( \mathbf{I}_{\text{in}} \in \mathbb{R}^{H \times W} \), a demosaic method \(\mathit{\mathbf{F}}\) aims to reconstruct \( \mathbf{I}_{\text{in}} \) into a RGB result \( \mathbf{I}_{\text{out}} \in \mathbb{R}^{H \times W \times 3} \).
We define the reconstruction task using the following formula:

\begin{equation}
  \mathbf{I}_{\text{out}} = \mathit{\mathbf{F}}\left(\mathbf{I}_{\text{in}}\right).
\end{equation}

\begin{figure}[!ht]
\centering
\includegraphics[width=0.48\textwidth]{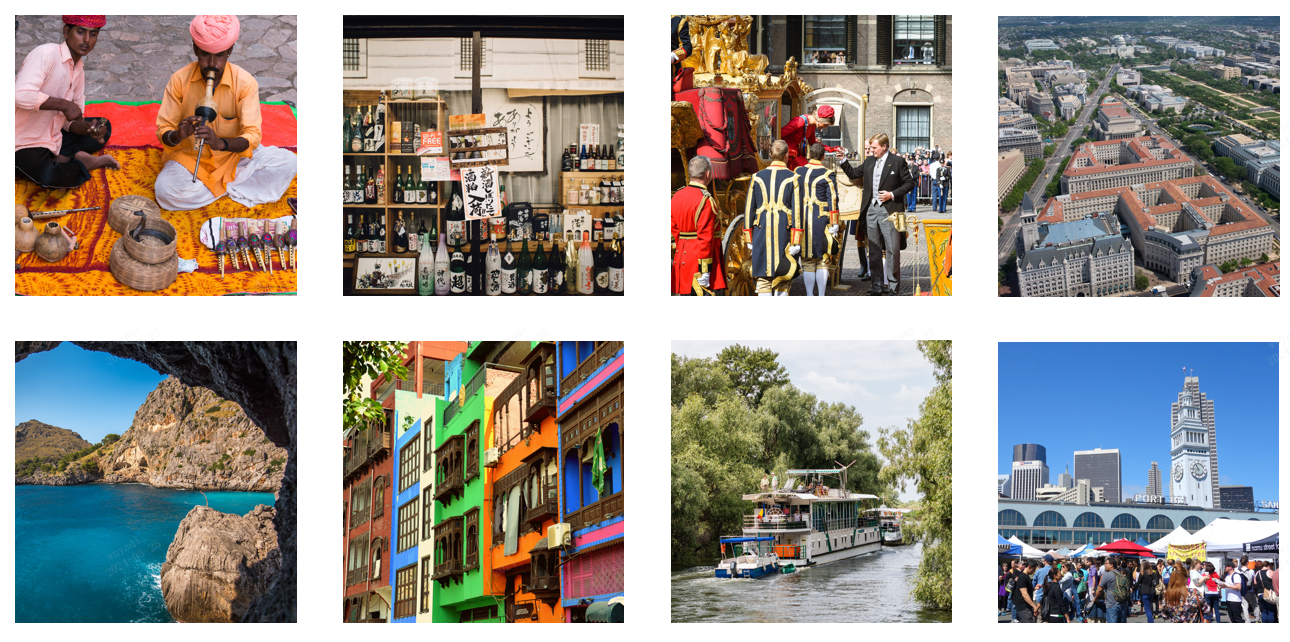}
\caption{Some sample images from the proposed training dataset. The scenes include natural landscapes, architectural views, and other such scenes.}
\label{fig:train_dataset}
\setlength{\belowcaptionskip}{0pt plus 3pt minus 2pt}
\end{figure}

\begin{figure}[t]
\centering
\includegraphics[width=0.48\textwidth]{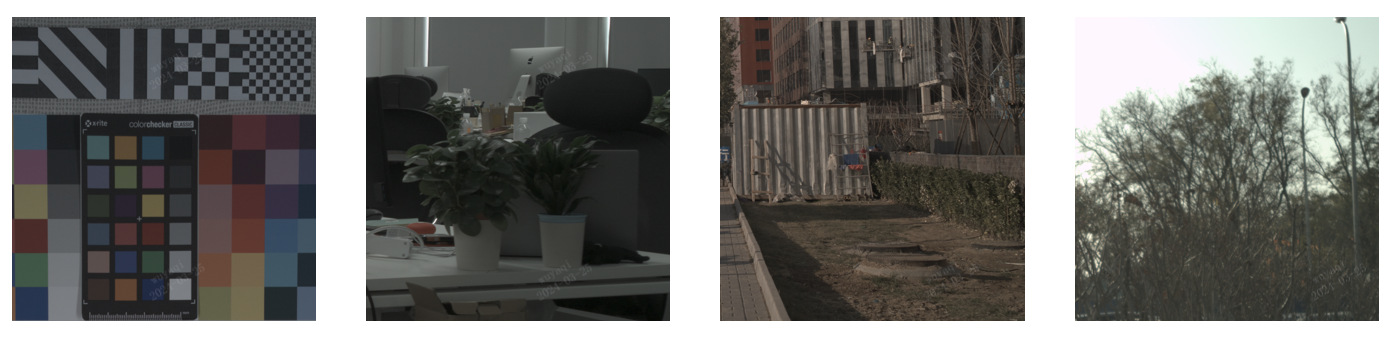}
\caption{Some sample images from the testing dataset. It includes various scenarios like indoor scenes and outdoor scenes}
\label{fig:test_dataset}
\setlength{\belowcaptionskip}{0pt plus 3pt minus 2pt}
\end{figure}

\subsection{Datasets}

As shown in Figure~\ref{fig:train_dataset}, the training dataset consists of 800 pairs of HybridEVS's input data and label results with a resolution of 2K. Both the input and label have the same spatial resolution. The input is of 10 bits in the .bin format, while the output is of 8 bits in the .png format. The validation and testing set has 50 scenes each.

During the testing phase, in order to achieve a more accurate and comprehensive evaluation, the challenge employed a testing dataset comprising both simulated and real-world scenarios. 
For real-world scenarios, we utilize commercially available smartphones equipped with imaging sensors such as Samsung's GN2 sensor to capture images. 
As illustrated in Figure~\ref{fig:test_dataset}, the scenes we captured encompass a variety of settings, including indoor image quality testing scenes, outdoor architectural scenes, outdoor strong lighting scenes, etc.

\subsection{Evaluation}
In this competition, we compare the recovered images with the ground-truth images.
We utilize the widely adopted Peak Signal-to-Noise Ratio (PSNR) and the complementary Structural Similarity (SSIM)~\cite{ssim} index to evaluate the quality of recovered images. 
Participants can view these metrics of their submission to further optimize the model's performance.

\subsection{Challenge Phase}
The challenge consisted of three phases as follows:
\begin{enumerate}
    \item Development: The registered participants get access to the data and baseline code, and can adopt them to train the models and evaluate their running time locally.
    \item Validation: The participants can upload their models to the remote server to check the fidelity scores on the validation dataset, and to compare their results on the validation leaderboard.
    \item Testing: The participants submit their final results, code, models, and factsheets.
\end{enumerate}

%% file: sec/3_result.tex
\section{Challenge Results}

Among the $108$ registered participants, $7$ teams successfully submitted their results, code, and factsheets in the final test phase. 
In order to ensure fairness in the competition, we have decided to exclude public datasets such as div2k from the final testing rankings and instead utilize remaining non-public datasets for the final ranking.
Table \ref{tab:result} reports the final test results and rankings of the teams. 
%

%
Finally, the USTC604 team clinched the first place in this challenge, followed by the lolers team in second place, and the Lumos Demosaicker team in third place.
The overall performance of all participating teams' solutions consistently exceeds 40 dB on the testing dataset, indicating that all participating teams can achieve relatively good reconstruction results. The PSNR of the first-place model reached 44.8464 dB, leading the second-place by 0.223 dB. The third-place model has a significant advantage over the fourth-place, with a margin of 0.5386 dB, which indicates the top three contestants exhibit considerable superiority.

\begin{table}[]
\caption{Results of MIPI 2024 challenge on the Demosaic for HybridEVS Camera. PSNR and SSIM are computed between the test results and ground truth. To ensure fairness in the competition, publicly available datasets such as DIV2K are excluded. The running time of input of $1080\times1920$ was measured. The measurement was taken on an NVIDIA Geforce GTX 1660Ti. \label{tab:result}}
\scalebox{0.8}{
\linespread{1.4} \selectfont
\begin{tabular}{c|c|ccc}
\hline
rank & team                  & PSNR               & SSIM              & \multicolumn{1}{l}{Time (s)} \\ \hline
1    & USTC604               & $\textbf{44.8464}_{(1)}$ & $\textbf{0.9854}_{(1)}$ & $51.315_{(6)}$                            \\
2    & lolers                & $44.6234_{(2)}$          & $0.9847_{(2)}$          & $18.231_{(2)}$                            \\
3    & Lumos\_Demosaicker    & $44.4951_{(3)}$           & $0.9845_{(3)}$          & $26.284_{(4)}$                            \\
4    & High\_speed\_Machines & $43.9564_{(4)}$          & $0.9838_{(4)}$          & $101.768_{(7)}$                           \\
5    & Yunfan                & $42.6508_{(5)}$          & $0.9810_{(5)}$          & $37.508_{(5)}$                            \\
6    & HIT-CVLAB             & $41.3280_{(6)}$           & $0.9780_{(6)}$          & $25.421_{(3)}$                            \\
7    & CougerAI              & $41.0736_{(7)}$          & $0.9752_{(7)}$          & $\textbf{6.331}_{(1)}$                    \\ \hline
\end{tabular}}
\end{table}

%% file: sec/4_methods.tex
\section{Methods}
\label{sec:methods}

\begin{figure}
  \centering
  \begin{subfigure}{\linewidth}
    \includegraphics[width=\linewidth]{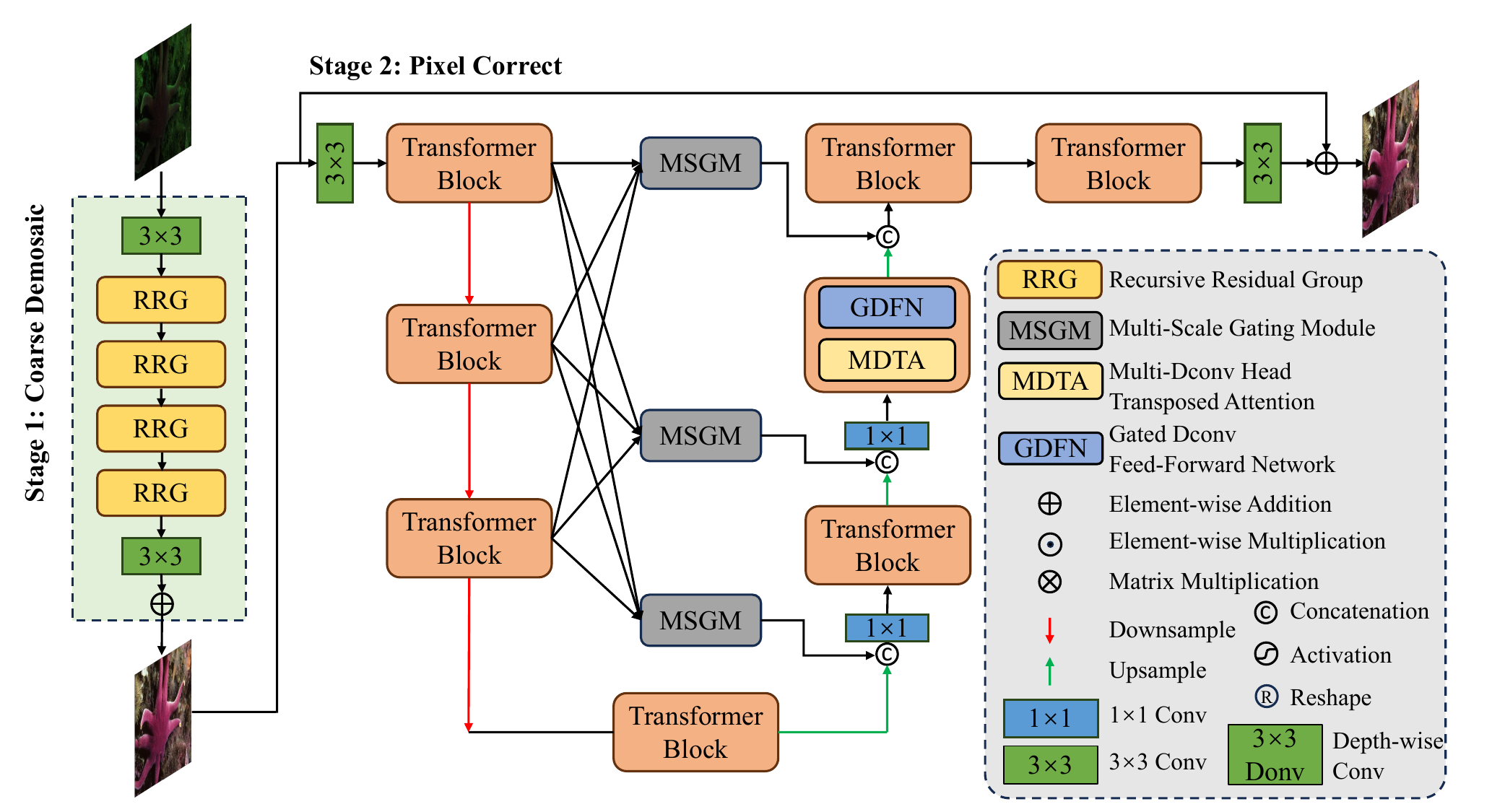}
    \caption{The architecture of DemosaicFormer.}
    \label{fig:short-a}
  \end{subfigure}

  \vspace{0.5cm}
  \begin{subfigure}{\linewidth}
    \includegraphics[width=\linewidth]{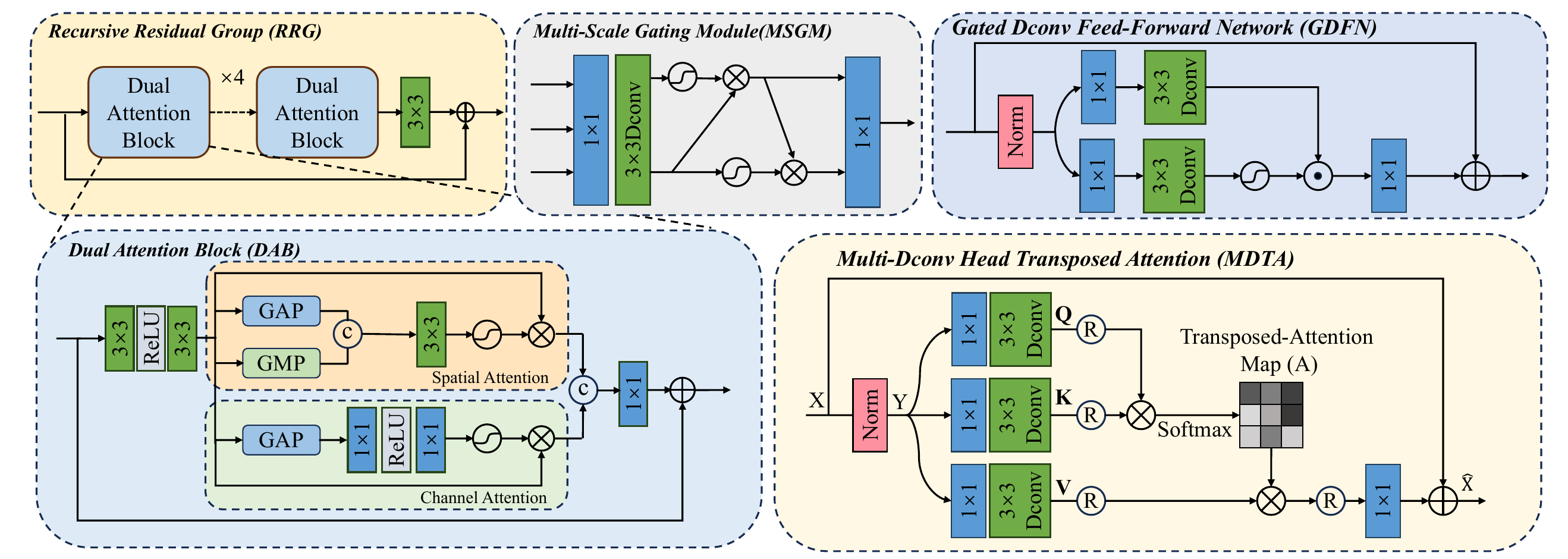}
    \caption{The structures of sub-modules in the main architecture.}
    \label{fig:short-b}
  \end{subfigure}
  \caption{The architecture of DemosaicFormer proposed by team USTC604 to demosaic the raw data captured by HybridEVS cameras.}
  \label{fig:short}
\end{figure}

\paragraph{\bf USTC604}

This team proposes a coarse-to-fine framework named DemosaicFormer which comprises a coarse demosaicing network and a pixel correction network (see Figure~\ref{fig:short}). For the coarse demosaicing stage, to produce a preliminary high-quality estimate of the RGB image from the HybridEVS raw data, this team introduces Recursive Residual Group (RRG) \cite{zamir2020cycleisp} which employs multiple Dual Attention Blocks (DABs) to refine the feature representation progressively.  For the pixel correction stage, aiming to enhance the performance of image restoration and mitigate the impact of defective pixels, this team introduces the Transformer Block\cite{zamir2022restormer} which consists of Multi-Dconv Head Transposed Attention (MDTA) and Gated-Dconv Feed-Forward Network (GDFN). The key innovation is the design of a novel Multi-Scale Gating Module (MSGM) applying the integration of cross-scale features inspired by \cite{chen2022simple}, which allows feature information to flow between different scales. Due to the inability to accurately model defective pixels, inspired by \cite{zamir2020cycleisp}, this team extracted the defect pixels map from the training data of the challenge to generate more diverse and realistic inputs for data augmentation. During the training phase, this team randomly rotated and flipped ground-truth images of training split, then sampled them according to the HybridEVS pattern, and randomly covered the sampled images with a defect pixels map. The augmentation technology is applied at the initial training of the proposed approach for improving the model's generalization and robustness. The training phase of the proposed method is divided into two stages:

\noindent \textbf{(1) Initial training of DemosaicFormer}. This team used a progressive training strategy at first. Start training with patch size $80\times80$ and batch size 84 for 58K iterations.  The patch size and batch size pairs are updated to $[(128^2,30), (160^2,18), (192^2,12)]$ at iterations [ 36K, 24K, 24K]. The initial learning rate is $5 \times 10^{-4}$ and remains unchanged when the patch size is 80. Later the learning rate changes with the Cosine Annealing scheme to $1 \times 10^{-7}$. The best model at this stage is used as the initialization of the second stage.

\noindent \textbf{(2) Fine-tuning DemosaicFormer}. This team starts training with patch size  $192 \times 192$ and batch size 12. The initial learning rate is $1 \times 10^{-4}$ and changes with Cosine Annealing scheme to $1 \times 10^{-7}$, including 20K iterations in total. Note that use the entire training data from the challenge without any data augmentation technologies at this stage. Exponential Moving Average (EMA) is applied for the dynamic adjustment of model parameters.

\begin{figure}[!ht]
    \centering
    \includegraphics[width=.95\linewidth]{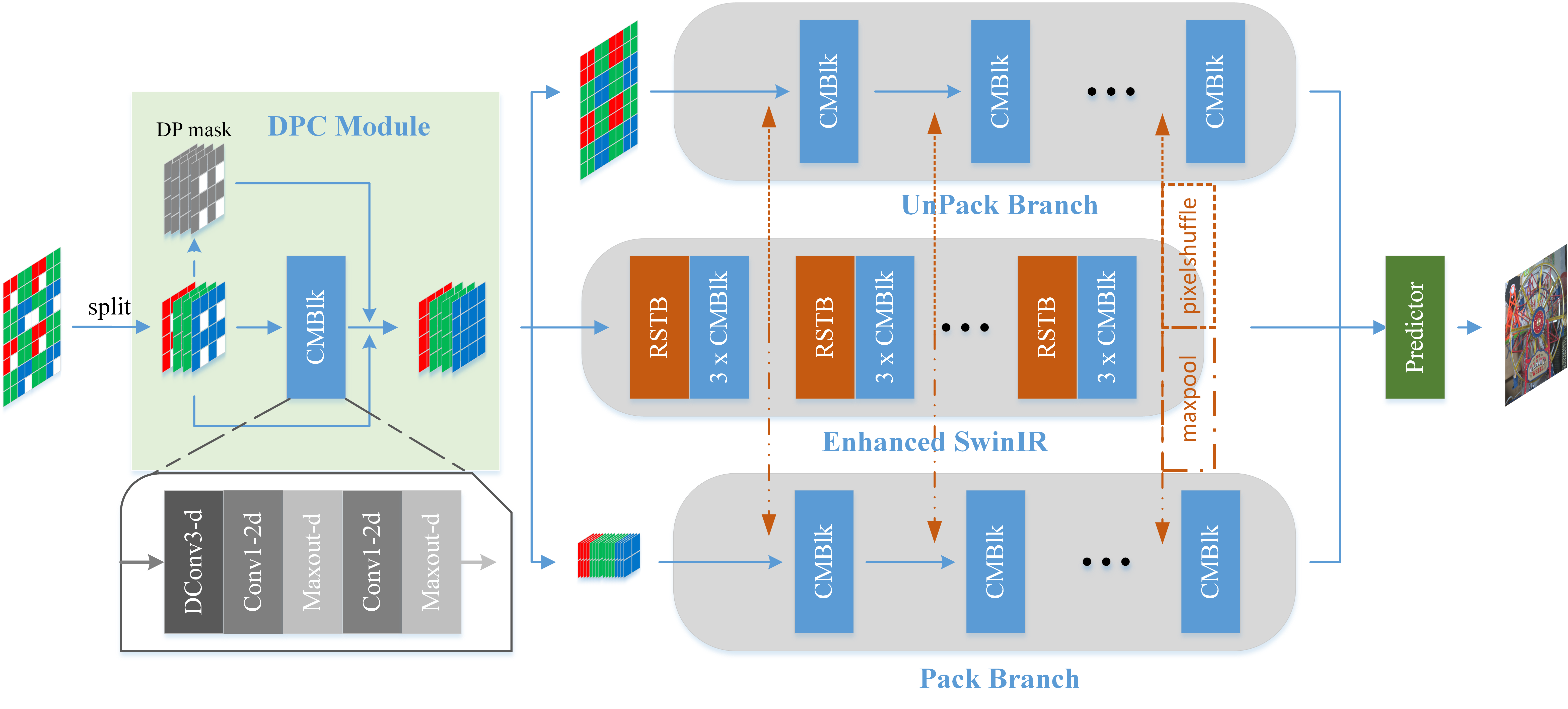}
    \caption{The multi-branch network architecture proposed by lolers team}
    \label{fig:lolers}
    \vspace{-7pt}
\end{figure}

\paragraph{\bf lolers}
This team proposed a SwinIR-based multi-branch network for HybridEVS demosaicing (see Figure~\ref{fig:lolers}). The backbone uses an enhanced SwinIR \cite{liang2021swinir} to extract rich features with long-range dependencies and pass them to the UnPack branch and Pack branch, which are used to indicate the spatial position of full resolution and provide uniform sampling, respectively. In addition, the authors propose a lightweight Cascaded Maxout Block (CMBlk), which consists of a depthwise convolution and multiple consecutive Maxouts, to give the model a powerful representation with a small number of parameters. The UnPack and Pack branches are composed of the same number of CMBlk.

For a defected QCFA raw, the authors first separate them into 4 single-channel inputs \emph{I}, and use a defect pixel mask (DP mask) to perform defect pixel correction (DPC). \[ DP\_mask = (I==0) \] \[ I=CMBlk(I)*DP\_mask+I \] After passing the DPC module, the 4-channel input \emph{I} is packed and unpacked respectively, restored to QCFA raw and 16-channel sub-images, and then input to the backbone and two branches. For the backbone network, it uses a SwinIR with a depth of 6 and follows 3 CMBlk after each RSTB to improve its feature extraction capability. Accordingly, the features of each layer are passed separately into two lightweight branches for feature enhancement. Finally, the three branches are fused to infer the final result.

In the training stage, the authors use the Adam optimizer, and the initial learning rate is set to 2e-4. First, they trained the DPC module and backbone by L1 loss. When the network is fitted, they fix the backbone's parameters and add two branches to continue training. After 50w iterations, the parameters of all three branches are updated in an end-to-end manner.

\begin{figure}[!ht]
    \centering
    \includegraphics[width=.95\linewidth]{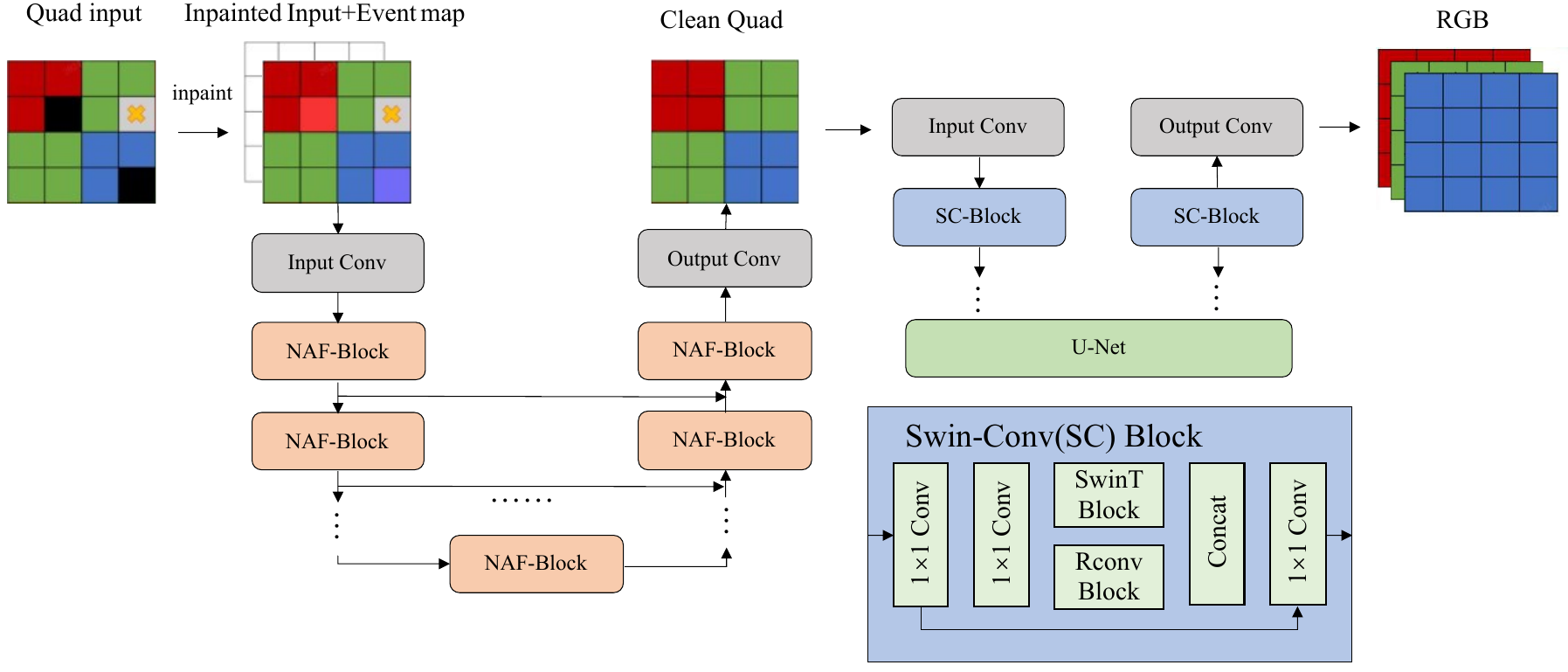}
    \caption{The network architecture of the proposed TSIDN}
    \label{fig:TSIDN}
    \vspace{-7pt}
\end{figure}

\paragraph{\bf Lumos Demosaicker}
This team introduces a novel two-stage network, termed the Two-Stage joint Inpainting and Demosaicing Network (TSIDN), as depicted in Figure~\ref{fig:TSIDN}. Initially, the network addresses the influence of event points by employing an inpainting process, which replaces them with the average values of neighboring pixels. Subsequently, the primary task is segmented into two stages, facilitating independent and joint training for each sub-network. 
The first stage features a Quad-to-Quad (Q2Q) network, which takes inpainted Quad Bayer data and event pixel maps as input. It utilizes a NAFNet \cite{chen2022simple} to effectively restore both event and defect pixels, integrating positional information to enhance restoration accuracy. Building upon this foundation, the second stage employs a Quad-to-RGB (Q2R) network based on SCUNet \cite{zhang2023practical} to focus on demosaicing. This network benefits from Swin-Conv and U-Net structures, ensuring efficient demosaicing performance.
During the training phases, strategies such as phase-based training and progressive learning are incorporated to enhance network performance. In the joint training stage, a progressive learning strategy is employed, starting with a patch size of 256 pixels, which progressively increases to 382 and then to 500 pixels. $l_1$ loss is utilized during the pretraining stage, while PSNR loss is applied during joint training. The initial learning rate is set at \(1\times10^{-4}\) and is gradually decreased to \(1\times10^{-7}\), contributing to the robustness and efficiency of the proposed network.

\paragraph{\bf High-speed Machines}

\begin{figure}[!ht]
\setlength{\abovecaptionskip}{0.cm}
\setlength{\belowcaptionskip}{-0.cm}
\centering
\includegraphics[width=0.47\textwidth]{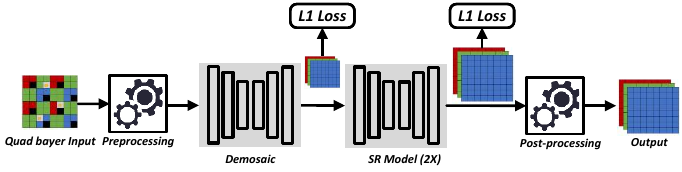}
\caption{Step-by-step Demosaic model for Hybrid Evs Camera(SBSDM).}
\label{fig:model}\textbf{}
\end{figure}

The Demosaic for HybridEVS Camera requires a 4$\times$ scale expansion, which is quite challenging for the model (see Figure~\ref{fig:model}). Therefore, a large number of parameters are needed for the model to learn, which in turn requires a substantial amount of data for training. However, the amount of training data provided by the competition is quite low. To address this difficulty, this team proposes the Step-by-step Demosaic model for HybridEVS Camera (SBSDM).

This team's model, SBSDM, is a two-stage model. The first stage involves transforming the Quad Bayer input into a $2\times$ RGB output image. In the data preprocessing part, since the PD point locations do not contain valid pixel information, this team has removed this information from the input and decomposed the original image into 14 channels. In the model, this team adopted the SPAN~\cite{wan2023swift} model and expanded the channels to 96. The second stage involves a $2\times$ super-resolution. Since there are many public models for super-resolution tasks, this team can conveniently use models pre-trained on large datasets as the second-stage model. Here, this team used EDSR~\cite{Lim_2017_CVPR_Workshops}, HAT~\cite{chen2023activating}, and SwinIR~\cite{liang2021swinir} to construct different models. Finally, this team trained the entire model end-to-end using the L1 loss function.

\begin{figure}[!t]
    \centering
    \includegraphics[width=1\linewidth]{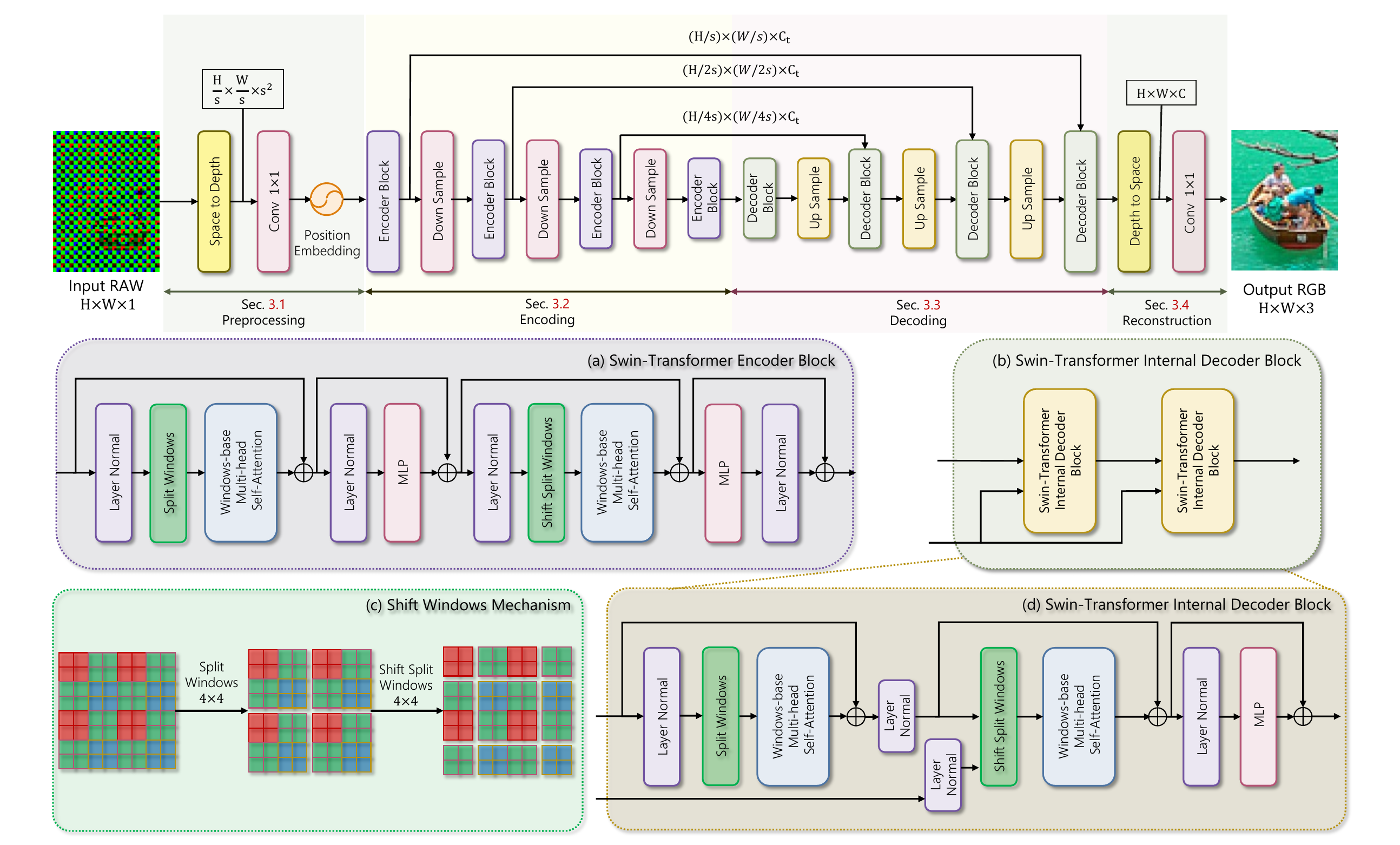}
    \caption{Overview of the proposed demosaicing method for event cameras: The process begins with preprocessing the RAW image, followed by feature extraction via an encoder using Swin Transformer blocks and a Shifted Window mechanism. The decoder, mirroring the encoder and including skip connections, reconstructs spatial details. The final stage is image reconstruction to produce the RGB output. Illustrated components include (a) the encoder architecture, (b) the Shifted Window mechanism for enhanced interaction, and (c) the decoder architecture.}
    \label{fig:yunfan:framework}
\end{figure}

\paragraph{\bf Yunfan}
This team has devised a comprehensive method for demosaicing images from event cameras by leveraging a sophisticated framework that combines the Swin Transformer~\cite{liu2022video} and U-Net~\cite{siddique2021u} architectures, as shown in Figure~\ref{fig:yunfan:framework}.
They have methodically outlined a multi-faceted approach consisting of preprocessing, encoding, decoding, reconstruction, and a novel loss function, each contributing uniquely to the image reconstruction process.

In the preprocessing phase, the team effectively transforms the input RAW image and reduces computational complexity using space-to-depth operations~\cite{hsu2024yolo} and $1\times1$ convolutions. The encoding stage, utilizing Swin Transformer blocks, extracts multi-scale features and captures long-range dependencies, while the decoding phase mirrors the encoding structure, progressively recovering the image's spatial resolution. 
The team's reconstruction module is adept at generating the final RGB image from upsampled features by reversing the preprocessing operations and refining the high-dimensional features into a standard color space. 

Notably, the team's strategic innovation lies in their two-stage training methodology that employs a Charbonnier loss for initial training and a Pixel Focus Loss for fine-tuning. They have meticulously engineered the Pixel Focus Loss to address long-tail distribution issues in training loss, focusing on edge detail enhancement. This loss function is instrumental in the model's ability to distinguish between high-frequency edge information and low-frequency color block differences, enhancing fine details' restoration.

Through this systematic approach, the team has strengthened the network's capability to learn global color distributions and local edge details, culminating in high-quality RGB image reconstruction. Their experiments demonstrate the efficacy of their method.

\begin{figure}[!t]
    \centering
    \includegraphics[width=1\linewidth]{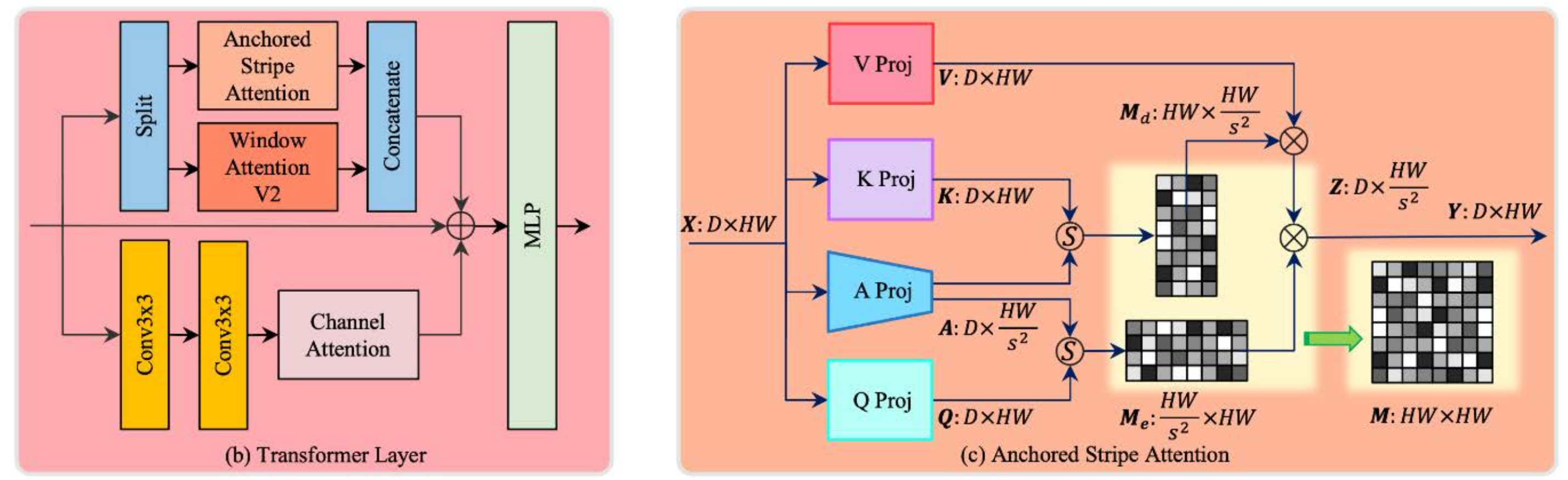}
    \caption{(b) The architecture of the adopted GRL layer. (c) The key anchored stripe attention mechanism. }
    \label{fig:GRL_Layer}
\end{figure}
\paragraph{\bf HIT-CVLAB}
This team designs a UNet structure network based on the GRL~\cite{li2023efficient} layer (see Figure~\ref{fig:GRL_Layer}), which explicitly models image hierarchies at global, regional, and local scales through anchored strip self-attention, window self-attention, and channel attention-augmented convolutions.

For training, the authors adopt mini-batch stochastic gradient descent (SGD) with a batch size of 64 for 600 epochs. 
The initial value of the learning rate is 3e-4 and gradually decreases to 1e-6 with a CosineAnnealing schedule. 
This learning rate decreasing strategy helps the model adjust parameters more carefully when it is close to convergence, thereby improving the generalization performance of the model and being able to better cope with Noise and uncertainty in training data.
Training loss functions include L1 loss, L2 loss, and Sobel loss. 

\paragraph{\bf CougerAI}

\begin{figure}[!t]
    \centering
    \includegraphics[width=1\linewidth]{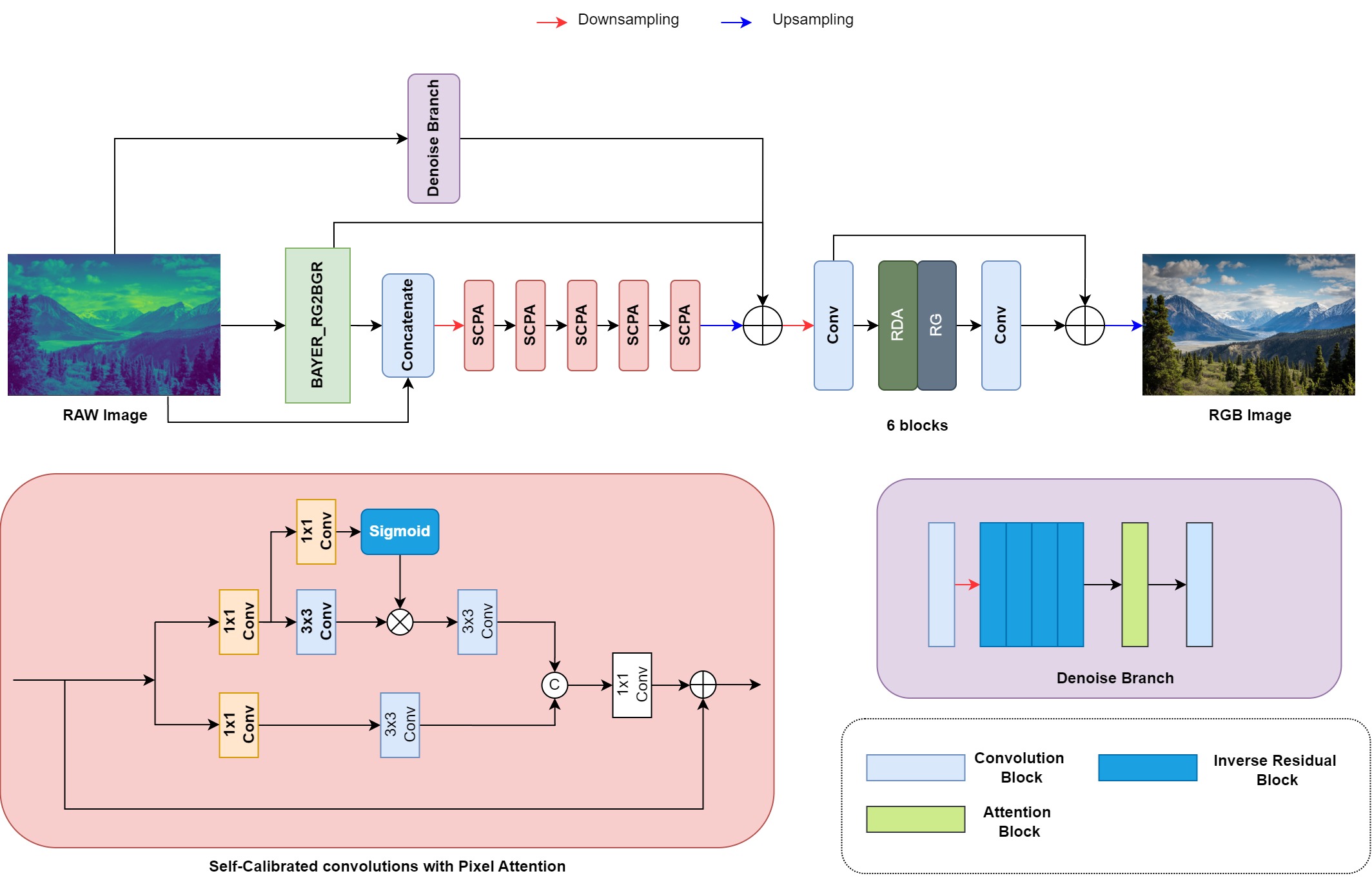}
    \caption{Multi-Stage Fusion Demosaicing architecture proposed by CougerAI team}
    \label{fig:Couger_AI}
\end{figure}

The demosaicing model begins with the initial step of converting the input image from its raw Bayer pattern (RGB) to the standard RGB color space using a BayerRG2BGR method (see Figure~\ref{fig:Couger_AI}). The resulting RGB image is then fused with the raw input image and jointly processed through a novel architecture inspired by recent advancements in computer vision.

This joint processing involves passing the concatenated images through a Self-Calibrated convolution with a pixel attention block (SCPA)\cite{Zhang,Kansal2023}, which enhances the spatial and spectral coherence of the input. Simultaneously, the raw input image is fed into a sophisticated denoising block\cite{Vasluianu} to produce a three-channel denoised image. The denoising block is composed of four inverse convolutional layers followed by an attention mechanism that effectively suppresses noise while preserving important image details.

After denoising, the denoised image and the output of the SCPA block are combined with the converted RGB image and fed through a downsampling layer to reduce computational complexity and enhance feature extraction. The downsampled features are then input to a residual learning block, which consists of a series of residual group blocks\cite{Egiazarian} followed by a residual channel dense attention block. This architecture allows the model to effectively capture both local and global contextual information, improving its ability to reconstruct the image.

Finally, the output of the residual learning block is added back to the input image and upsampled to produce the final demosaiced image. 
 
%% file: sec/5_conclusions.tex
\section{Conclusions}
In this report, we review and summarize the methods and results of MIPI 2024
challenge on the Demosaic for HybridEVS Camera. The participants have made significant contributions to this challenging track, and we express our gratitude for the dedication of each participant.

%% file: sec/6_teams.tex
\section{Teams and Affiliations}
\label{append:teams}

\subsection*{\bf USTC604}
\noindent
{\bf Title:} DemosaicFormer: Coarse-to-Fine Demosaicing Network for HybridEVS Camera \\
{\bf Members:}\\
Senyan Xu$^1$ (\href{syxu@mail.ustc.edu.cn}{syxu@mail.ustc.edu.cn})\\
 Zhijing Sun$^1$\quad Jiaying Zhu$^1$\quad Yurui Zhu$^1$\quad  Xueyang Fu $^1$\quad  Zheng-Jun Zha$^1$\\
{\bf Affiliations:}\\
$^1$ University of Science and Technology of China 

\subsection*{\bf lolers}
\noindent
{\bf Title:} Multi-Resolution SwinMaxIR for QCFA Raw Demosaic \\
{\bf Members:}\\
Jun Cao$^1$ (\href{caojun6@xiaomi.com}{caojun6@xiaomi.com})\\
 Cheng Li$^1$\quad Shu Chen$^1$\quad  Liang Ma $^1$\\
{\bf Affiliations:}\\
$^1$ Xiaomi Inc., China

\subsection*{\bf Lumos Demosaicker}
\noindent
{\bf Title:} Two-Stage joint Inpainting and Demosaicing Network\\
{\bf Members:}\\
Shiyang Zhou$^1$ (\href{shiyangzhou2023@163.com}{shiyangzhou2023@163.com})\\
Haijin Zeng$^2$\quad Kai Feng$^3$\quad Yongyong Chen$^1$\quad Jingyong Su$^1$\\
{\bf Affiliations:}\\
$^1$ Harbin Institute of Technology (Shenzhen)\\
$^2$ IMEC-UGent\\
$^3$ Northwestern Polytechnical University

\subsection*{\bf High-speed Machines}
\noindent
{\bf Title:} Step-by-step Demosaic model for Hybrid Evs Camera\\
{\bf Members:}\\
Xianyu Guan$^1$ (\href{guanxianyu@xiaomi.com}{guanxianyu@xiaomi.com})\\
Hongyuan Yu$^1$\quad Cheng Wan$^3$\quad Jiamin Lin$^1$\quad Binnan Han$^1$\quad Yajun Zou$^1$\quad Zhuoyuan Wu$^1$\quad Yuan Huang$^1$\quad Yongsheng Yu$^2$\quad Daoan Zhang$^2$\quad Jizhe Li$^1$\quad Xuanwu Yin$^1$\quad Kunlong Zuo$^1$\\
{\bf Affiliations:}\\
$^1$ Multimedia Department, Xiaomi Inc.\\
$^2$ University of Rochester \\
$^3$ Georgia Institute of Technology

\subsection*{\bf Yunfan}
\noindent
{\bf Title:} Event Camera Demosaicing via Swin Transformer and Pixel-focus Loss\\
{\bf Members:}\\
Yunfan LU$^1$ (\href{ylu066@connect.hkust-gz.edu.cn}{ylu066@connect.hkust-gz.edu.cn}) \\
Yijie XU$^1$ Wenzong MA$^1$ Weiyu GUO$^1$ Hui XIONG$^1$\\
{\bf Affiliations:}\\
$^1$ AI Thrust, The Hong Kong University of Science and Technology (Guangzhou)

\subsection*{\bf HIT-CVLAB}
\noindent
{\bf Title:} Efficient and Explicit Hierarchies Modelling Network for HybridEVS Camera Demosaic \\
{\bf Members:}\\
Wei Yu$^1$ (\href{20b903014@stu.hit.edu.cn}{20b903014@stu.hit.edu.cn})\\
 Bingchun Luo$^1$\\
{\bf Affiliations:}\\
$^1$ Harbin Institute of Technology

\subsection*{\bf CougerAI}
\noindent
{\bf Title:} Multi-Stage Fusion Demosaicing with Integrated Pixel Attention and Residual Learning \\
{\bf Members:}\\
Sabari Nathan$^1$ (\href{sabari@couger.co.jp}{sabari@couger.co.jp})\\
Priya Kansal$^1$\\
{\bf Affiliations:}\\
$^1$ Couger Inc, Japan
